\documentstyle[12pt]{article}

\def\be{\begin{equation}}
\def\ee{\end{equation}}
\def\br{\begin{eqnarray}}
\def\er{\end{eqnarray}}
\def\nonu{\nonumber}
\def\o{\over}
\def\cl{{\cal L}}
\def\pa{\partial}
\def\u2{\mid u\mid^2}
\def\vp{\varphi}
\def\nonu{\nonumber}
\def\br{\begin{eqnarray}}
\def\er{\end{eqnarray}}
\def\be{\begin{equation}}
\def\ee{\end{equation}}

\def\({\left(}
\def\){\right)}

\newcommand{\ct}[1]{\cite{#1}}
\newcommand{\bi}[1]{\bibitem{#1}}

\relax

\def\o{\over}

\def\pa{\partial}

\def\ra{\rightarrow}

\def\tp0{\Theta_{+}^{(0)}}
\def\tm0{\Theta_{-}^{(0)}}

\def\u2{\mid u\mid^2}
\def\z2{\mid z\mid^2}

\def\vp{\varphi}

\def\cl{{\cal L}}

\def\one{\hbox{{1}\kern-.25em\hbox{l}}}
\def\0#1{\relax\ifmmode\mathaccent''7017{#1}%
        \else\accent23#1\relax\fi}

\def\PRL#1#2#3{{\sl Phys. Rev. Lett.} {\bf#1} (#2) #3}
\def\NPB#1#2#3{{\sl Nucl. Phys.} {\bf B#1} (#2) #3}

\def\PLB#1#2#3{{\sl Phys. Lett.} {\bf #1B} (#2) #3}
\def\JMP#1#2#3{{\sl J. Math. Phys.} {\bf #1} (#2) #3}

\begin{document}

\begin{titlepage}
\vspace*{-2 cm}
\noindent

\vskip 3cm
\begin{center}
{\Large\bf Ring-shaped exact Hopf solitons }
\vglue 1  true cm
C. Adam$^1$ and  J. S\'anchez-Guill\'en$^2$
\vspace{1 cm}

$^1${\footnotesize Wolfgang Pauli Institut
 c/o Institut f\"ur Mathematik,\\
Universit\"at Wien,\\
 1090 Vienna, Austria}

\vspace{1 cm}
$^2$ {\footnotesize Departamento de F\'\i sica de Part\'\i culas,\\
Facultad de F\'\i sica\\
Universidad de Santiago\\
E-15706 Santiago de Compostela, Spain}

\medskip
\end{center}

\normalsize
\vskip 0.2cm

\begin{abstract}

{The existence of ring-like structures in exact hopfion solutions
  is shown.}

\end{abstract}

\vskip 2cm

\end{titlepage}
\section{Introduction }

Extended solutions are of central importance in different applications of
modern field theory, from high energy  to condensed matter physics.
Their
relevance is parallel to the  difficulty in the analysis of the nonlinear
theories which encode them and the scarce exact results in higher
dimensions.
To circumvent one of the main problems, the scaling instablity for scalar
fields beyond one spatial dimension
found by Skyrme \cite{skyrme} and
formalized by Derrick \cite{Derr}, one obvious
possibility  is nonpolynomial lagrangians, which are not unfamiliar. Among
those attempts the work by Deser, Duff and
Isham \ct{Deser},
where the simplest choice of just the
power required to balance the scaling is analyzed, is of special interest.
As the authors discuss, such models have, of
course,
no free field expansion around a trivial vacuum.  But they can have a
semiclassical formulation around
 nontrivial solutions and
eventually a small time dependent  perturbation of the static solutions.
One solution was found and
 the importance of global transformations was emphasized. This
  model was further extended by Nicole \cite {Nico}
and  put in a more general framework by Kundu \cite{Ku}. It  was
rediscovered independently
in a series of papers \cite{afz} in the context of a new proposal for a
generalized  integrability  \cite{noso}, finding infinitely many analytic
solutions with  general Hopf indices. This was an unsolved problem, which
is important for soliton physics and because of  the many applications of
those  maps, combining
topology and geometry. In their last  paper \ct{baf}, a new  feature of the
Hopfion solutions was discovered, namely, a line singularity in an
infinitely thin
tube along the z-axis from
 a special current
with a non-conserved charge, provided by the geometric  method.

In this Letter we extend  the analysis to find that there is in fact
another solution
of the ring type, which is interesting, as ring structures  are typical  in
higher dimensional soliton analysis, both in  numerical \ct{sut} and analytical
approximations \ct{ward}. In fact, these currents and solutions should
be relevant  for the analytical study of soliton scattering. We pay also
special
attention to the symmetries, including global aspects of the solution,
 as many aspects of the model are generic to other
local formulations of topological degrees of freedom. Time dependence
analysis is also a natural possiblity in the the Generalized Zero Curvature
approach, as it preserves Lorentz  covariance.

\section{The ring-like solutions}

As shown in \cite{baf}, the only models involving the antisymmetric tensor
in the complex field $u$
\be
h_{\mu\nu} = - i \, ( \pa_{\mu}u \pa_{\nu}u^* -
\pa_{\nu}u \pa_{\mu}u^* )
\label{hmnu}
\ee
which satisfy without constraints the integrability criterion of  the
geometric approach of \cite{noso}
(i.e. infinitely many conserved currents) 
and which can be derived from an action
principle, are those where
the lagrangian is a functional of $h^{2}/f^{2}$, where $f$ is a real
function of $u$ and $u^*$.
If one further asks for scaling invariance to allow for  stable static
solutions, we are lead to the class of
models given by the Lagrangian
density
\be \label{lagr}
\cl \equiv  \left( {h_{\mu\nu}^2\o 2f^2}
\right)^{3\o 4}
\label{nicemodel}
\ee
which generalizes and explains choices and  solutions found before  \ct
{Deser} - \ct {afz}.

We assume from now
on that $f$ only depends on $uu^*$.
If $\lim_{|\vec x|\to \infty}
u(\vec x)=u_0 ={\rm const}$ is assumed, then
the domain space $I\hspace{-0.1cm}R^3$ has the topology of the three-sphere
$S^3$. If, in addition, the target space may be identified with the
two-sphere $S^2$ via stereographic projection, then
$u$ may be interpreted as a map
 $S^3 \to S^2$, which is characterized by an integer winding number
(the Hopf index). In this case there exists an infinite number of static,
soliton-like
solutions to the equations of motion of the model, and these all have
integer Hopf index.  These solutions were first found in \cite{afz}
and  they are obtained by inserting
into the static equations of motion
\be
-h_{ij}\pa^j u \pa^i h^2 + 4 h^2 \pa^i h_{ij}\pa^j u +i (h^2 )^2 \pa_{u^*}
f =0
\ee
(where $h^2 \equiv h_{ij}h^{ij}$)
the product ansatz
\be
u ( \eta, \xi, \vp ) \equiv R (\eta) e^{ i  (m \xi+ n \vp)}
\label{genu}
\ee
in  toroidal
coordinates
\br
x &=&  q^{-1} \sinh \eta \cos \vp \;\;, \;\;
y =  q^{-1} \sinh \eta \sin \vp   \nonu \\
z &=&  q^{-1} \sin \xi \quad ;    q = \cosh \eta - \cos \xi .
\label{tordefs}
\er
 As explained elegantly in \cite {baf}, the Ansatz (\ref{genu}) follows
from the conformal symmetry of
the equations of motion.
If one assumes that $f$ is a function of
$T\equiv R^2 \equiv uu^*$ only,  $\pa_{u^*}f$ simplifies to
$\pa_{u^*}f =f_{,T}\, u$. With the Ansatz (\ref{genu}),
this results in an ordinary differential
equation for $R(\eta)$, which may be conveniently expressed in terms of
$T\equiv R^2 \equiv uu^*$ as
\be
\left( \ln \frac{T_{,\eta}}{f}\right) _{,\eta}
= \frac{\cosh \eta }{\sinh \eta}
\frac{n^2 -2m^2 \sinh^2 \eta}{n^2 +m^2 \sinh^2 \eta}
\ee
and $X_{,\eta}$ denotes derivative of $X$ w.r.t. $\eta$.
Further, we assume $m^2 >n^2$ in the sequel.
A first integral may be found easily,
\be \label{1-int}
\frac{T_{,\eta}}{f}=k_1 \frac{\sinh\eta}{(n^2 +m^2\sinh^2 \eta)^{3/2}}
\ee
(here $k_1$ is a constant of integration),
whereas for a further integration the explicit form of the function $f(T)$
is needed.

In the end, we shall choose $f=(1+T)^2$, because we are refering  to the
solutions of Babelon and Ferreira \cite{baf}, but let us briefly mention a
class of functions $f(T)$ that leads to a target space with the topology of the
two-sphere and, therefore, to genuine Hopf solitons, which can have useful
applications. For $f=(1+T)^2$,
the expression $h_{ij}/f$ in the Lagrangian density (\ref{lagr})
is, in fact, just the pull-back under the
map $u$ of the area two-form
\be \label{s2-ar}
d\Omega =-i\frac{dzdz^*}{(1+zz^*)^2}
\ee
on the two-sphere. A pull-back of this two-form under maps $S^2 \to S^2$
will lead to further acceptable area two-forms (i.e. area two-forms
respecting the topology of the target space). If we want to maintain the
simple dependence $f=f(T)$, then a class of allowed maps is
 \be
\phi \, :\;  z\to \sqrt{g(zz^*)}e^{il\arg (z)} .
\ee
These are indeed maps $S^2 \to S^2$ provided that $g(0)=0$ and $g(\infty)=
\infty$. Further, $l$ must be an integer. The pullback of the area two-form
(\ref{s2-ar}) is
\be
\phi_* (d\Omega ) =-i\frac{dzdz^*lg'}{(1+g)^2}
\ee
therefore any function $f$ of the type
\be
f(T)=\frac{(1+g(T))^2}{l\, g_{,T}}
\ee
leads to a theory (\ref{lagr}) with genuine Hopf solitons.

In the sequel we restrict to the simplest case $f=(1+T)^2$ (the area
two-form on the two-sphere).
Then the first integral (\ref{1-int}) may be easily integrated
to yield
\be
{1 \o 1+T } = { k_1 \o  (m^2 -n^2)} { \cosh \eta
  \o ( n^2  + m^2
\sinh^2 \eta )^{1/2} } +k_2
\label{fspone}
\ee
where $k_1 $ and $k_2$ are two constants of integration.
These constants have to be fixed by imposing some boundary conditions on
the field $u$. For this purpose let us introduce the unit vector $\vec n$
related to $u$ via stereographic projection
\be
{\vec n} = {1\o {1+\mid u\mid^2}} \, ( u+u^* , -i ( u-u^* ) , \u2 -1 ) \; ;
\qquad
u  = \frac{n_1 + i n_2}{1 - n_3}.
\label{stereo}
\ee
If $u$ is supposed to be a true Hopf map, then the number of allowed
boundary conditions is, in fact, very restricted.
The point is that a true Hopf map should cover the whole target $S^2$,
including the north pole ($\vec n = (0,0,1)$, or $T=\infty$) and the
south pole ($\vec n = (0,0,-1)$, or $T=0$). Therefore, the boundary
conditions should be chosen such that $T$ varies between $T=0$ and
$T=\infty$. Further, the pre-images of the north pole and the south pole
should be one-dimensional lines in $I\hspace{-0.1cm} R^3$. However,
the only values
of $\eta$ which define one-dimensional lines (instead of two-dimensional tori,
as is the general case), are $\eta =0$, which
defines the $z$ axis (together with spatial infinity), and $\eta =\infty$,
which defines the circle
\be
C=\{\vec x \in I\hspace{-0.12cm} R^3 :\;  z=0 \;
\wedge \; r^2 =1\}.
\ee
Therefore, there are two possible choices for the boundary conditions, namely
$T(\eta =0)=0$, $T(\eta =\infty) =\infty$, or
$T(\eta =0)=\infty$, $T(\eta =\infty) =0$.

In \cite{baf} the second option was chosen (which we call $T^{(2)}$ for
convenience),
\be
T^{(2)}=
 {   \cosh \eta  - \sqrt{n^2/m^2 +\sinh^2 \eta}  \o
 \sqrt{1 + m^2/n^2\,\sinh^2 \eta  }  -   \cosh \eta  } \quad ,
T^{(2)}(\eta =0)=\infty \; ,T^{(2)}(\eta =\infty) =0.
\ee
Let us now investigate our first option, for which we get
\be
T^{(1)}=
 (T^{(2)})^{-1} \quad ,
T^{(1)}(\eta =0)=0 \; ,T^{(1)}(\eta =\infty) =\infty .
\ee
Now one could simply identify the solitons with the position of their
singular value $|u|=\infty$, then the solitons of type $T^{(2)}$ would be
identified with the straight line $x=0,y=0$ (the $z$ axis), and the
solitons of type $T^{(1)}$ would be ascribed to the circle $C$.
However, this identification is in principle  quite arbitrary, and requires
a physical
motivation.

In \cite{baf} it was shown  that the solitions of type $T^{(2)}$ can
indeed be identified with the z axis in a well-defined manner. More
precisely, there exists a conserved current $ J_\mu$ which has the property
that for solitons of type $T^{(2)}$ it is singular along the $z$ axis.
Moreover, a constant flow of this current is emerging from the singular line
(the $z$ axis). Here we want to investigate the equivalent problem for
solitons of type $T^{(1)}$, which is relevant as rings are common in
higher dimensional solitons \ct{sut}, \ct{ward}.

So let us briefly review  and further develop
some results of \cite{baf} on the above-mentioned
conserved current, with some  more details required for our purposes.
There exists
 (among other symmetries) a symmetry of
the action under a transformation which is a combination of a dilatation
on three-dimensional domain space plus a specific
transformation on target space. The infinitesimal version of this
symmetry transformation
is given in \cite{baf}, but it is, in fact, not difficult to obtain the
transformation for finite transformation parameter. Under a dilatation
transformation $x\to \Lambda^3 x\equiv e^{3\lambda}x$ (the power three
of the dilatation parameter is chosen for later convenience) the action of
the theory scales like
\be
S\to \Lambda^{-3}S ,
\ee
therefore the theory is invariant if the dilatation is combined with
a transformation of the target space variable,  $u\ra v(u ,\bar u)$, such
that
\be \label{BF-eq}
\frac{dud\bar u}{(1+\bar u u)^2} \ra \frac{dvd\bar v}{(1+\bar vv)^2}
=\Lambda^2 \frac{dud\bar u}{(1+\bar u u)^2}.
\ee
If we introduce the real coordinates on target space $u=T^{1/2}
e^{i\phi}$ (angle and radius squared on the Euclidean plane) and assume that
$v=(\tilde T)^{1/2}(T)e^{i\phi}$ (i.e. $u$ and $v$ have the same argument,
and the modulus of $v$ is a function of the modulus $T$ only) then we get the
equation
\be
\frac{\tilde T '(T)dTd\phi}{(1+\tilde T)^2}=\Lambda^2 \frac{dTd\phi}{
(1+T)^2}
\ee
or
\be
\frac{\tilde T'}{(1+\tilde T)^2} =\frac{\Lambda^2}{(1+T)^2}
\ee
with the solution
\be \label{int-const}
\frac{1}{1+\tilde T}=\frac{\Lambda^2}{1+T} +c
\ee
where $c$ is a constant of integration. If we require the boundary condition
$\tilde T(0)=0$ then we get
\be
\tilde T=\frac{\Lambda^2 T}{\Lambda^2 +(1+T)(1-\Lambda^2)}
\ee
or
\be \label{TS-transf}
v=\frac{\Lambda u}{[\Lambda^2 +(1+\bar uu)(1-\Lambda^2)]^{1/2}}.
\ee
This $v$ indeed fulfills Eq. (\ref{BF-eq}) as may be checked easily.
In addition, it reduces to the transformation of Babelon and Ferreira
\cite{baf} for infinitesimal $\lambda$ (i.e. $\Lambda =1+\lambda$).
Therefore Eq. (\ref{TS-transf}) is the required symmetry transformation
on target space.\footnote{Observe that this transformation has the funny
property that it is
well-defined only for $\Lambda^2 \le 1$, i.e., for scaling
transformations which shrink distances .}

The conserved Noether current related to this symmetry transformation is
\be
J_\mu =x^\nu \Theta_{\mu\nu} +j_\mu
\ee
where $\Theta_{\mu\nu}$ is the canonical energy-momentum tensor of the
theory, and $j_\mu$ is
\be
j_\mu =-i\left( \frac{h^2}{2(1+T)^4}\right)^{-1/4}\frac{1}{2(1+T)^3}
h_{\mu\nu} (u\pa^\nu u^* - u^* \pa^\nu u) .
\ee
Here, the first term of $J_\mu$ (containing the energy-momentum tensor) is
due to the space dilatation, whereas the second term, $j_\mu$, is due to
the specific target space transformation (\ref{TS-transf}).
The current obeys the
conservation equation $\pa^\mu J_\mu =0$. For static configurations
$u (\vec x)$ this conservation equation may be used to derive the relation
\be \label{cons}
E\equiv \pa_t \int d^3 xJ^0 =\int_\Sigma d \vec f \cdot \vec J
\ee
where $E$ is the static energy of the static configuration,
\be \label{stat-e}
E=4\pi^2 \sqrt{|m||n|(|m| + |n|)}\, ,
\ee
and the integral in the r.h.s. of (\ref{cons}) is an integral over
surfaces $\Sigma$ which
surround the singularities of the current $\vec J$.
It turns out that the first term of the current for static fields,
$x^j \Theta_{ij}$, is regular everywhere and may, therefore, be ignored
in the surface integral of (\ref{cons}) provided that the integration
surfaces are chosen such that the enclosed volume is infinitesimal.
For the current $\vec j$ we find after some calculation
\be
\vec j = \left(  \left\vert
\frac{T_{,\eta}}{(1+T)^2}\right\vert \right)^{1/2} \frac{T}{1+T}
\left( \frac{n^2}{\sinh^2 \eta} +m^2\right)^{3/4} ( \cosh \eta -\cos \xi )^2
\vec e_\eta
\ee
where $\vec e_\eta = (\cosh\eta - \cos\xi )^{-1}\nabla\eta$ is a vector of
unit length which is perpendicular to the surfaces of constant $\eta$ (tori).
Observe that $\vec e_\eta $ is pointing into the interior of the tori,
because $\eta$ is growing in this direction.
We may use the first integral of the equations of motion, (\ref{1-int}),
for the first factor on the r.h.s.
containing $T_{,\eta}$, and find
\be
 \vec j = \sqrt{\vert m\vert \vert n\vert (\vert m \vert +
\vert n\vert )}\frac{T}{1+T}
\frac{(\cosh \eta -\cos \xi )^2}{\sinh \eta} \vec e_\eta .
\ee
If we ignore the factor $T/(1+T)$ for the moment, then we see that the
remaining expression is singular both for $\eta =0$ (along the $z$ axis)
and for $\eta =\infty$ (along the circle $C$). Depending on whether we choose
$T^{(1)}$ or $T^{(2)}$ for $T$, one of the two singularities gets cancelled,
whereas the other remains. For $T=T^{(2)}$ (the case which was studied in
\cite{baf}), the singularity along the $z$ axis remains. In this case we choose
a very large torus $\eta  <<1$ as integration surface. For the regular terms
the limit $\eta  \to 0$ may be performed, such that the integration is
extended to the whole space. For the singular $\vec j$ the surface integral
should be performed for a finite $\eta$ and the limit $\eta \to 0$ should
be taken afterwards.
The surface element on the torus surface $\eta ={\rm const}$ is
\be
d\vec f = \vec e_\eta \frac{\sinh \eta}{(\cosh\eta -\cos \xi )^2}d \vp d
\xi
 \ee
therefore the surface integral is
\br  \label{surf-i}
\int _{\eta ={\rm const}} \vec j\cdot d\vec f &=&
\sqrt{\vert m\vert \vert n\vert (\vert m \vert +
\vert n\vert )} \frac{T}{1+T}\int d\vp d\xi  \nonumber \\
&=&
4\pi^2 \sqrt{\vert m\vert \vert n\vert (\vert m \vert +
\vert n\vert )} \frac{T}{1+T}
 \er
 For $T=T^{(2)}$ this should be evaluated in the limit $ \eta  \to 0$ for
which $T^{(2)}/(1+T^{(2)})$ is equal to one. Therefore the total
flux emerging from the singular line (the $z$ axis) is
 \be \label{flux}
{ \rm flux} =4\pi^2 \sqrt{\vert m\vert \vert n\vert (\vert m \vert +
\vert n\vert )} =E
 \ee
where $E$ is the static energy (\ref{stat-e}).

On the other hand, for $T=T^{(1)}$ the singularity is located at the circle
$C$, therefore a
tiny torus (large  $ \eta$) should be excluded from the integration region.
So we take the limit $ \eta  \to  \infty$ now in the surface integral
(\ref{surf-i}). But $T^{(1)}/(1+T^{(1)})$ is equal to one in this limit,
so we find again for the flux the same previous expression  (\ref{flux}) .

Therefore, for solutions of the type $T^{(1)}$ the singular line is the circle
$C$ and a non-zero flux of total amount given in (\ref{flux}) emerges
from this singular circle. As a consequence,
the solutions of type $T^{(1)}$ are
characterized by a ring-like structure, where the ring is located at the
position of the circle $C$, which , as said, is relevant for their physics
and specially for the scattering.

\section{Discussion}

So we indeed found that in addition to the solutions of type $T^{(2)}$
originally obtained by Aratyn, Ferreira and Zimerman (AFZ) in \cite{afz},
which are characterized by a straight
line of singular flux according to Babelon and Ferreira, there exist
solutions where the singular flux is
located along the circle $C$, forming thereby a ring-like structure.
In hindsight, this result is not so surprising, and the ring-like structure
is, in fact, the generic case. To see this, let us invoke a further
symmetry of the model, namely  constant rotations of the target space
$S^2$. In domain space $I\hspace{-0.1cm}R^3$ such a rotation rotates
different level curves (i.e., curves of constant $u=u_0$ for different
values of $u_0$) into each other, because these level curves are the
pre-images of points of the target $S^2$ under the map $u$.
Generically, these level curves are circles, with the only exception of
the $z$ axis. Therefore, any rotation on target space which moves the
north pole and the south pole will transform a solution of type $T^{(2)}$
into a new solution where the line of singular flux is located along a
circle.

On the target space coordinate $u$ such rotations are represented by
modular transformations $u\to (a+bu)/(c+du)$, where $ad-bc =1$. A general
modular transformation is, however, not compatible with the simple Ansatz
(\ref{genu}) which was used by AFZ to find solutions. The only non-trivial
modular transformation compatible with the ansatz (\ref{genu}) is the
inversion map $u\to (1/u)$. And indeed, the composition of the inversion
map with the map $(m,n)\to (-m,-n)$ (which again maps a solution to
another solution with the same energy) precisely maps the AFZ solutions
$T^{(2)}$ to the solutions of type $T^{(1)}$ discussed in this paper.

Besides their interest for the scattering, the results can also be
useful - given
the generic features of the theory considered - for other higher
dimensional models  on the sphere
and/or using similar scaling arguments, like Skyrme theory and its restriction
to the $SU(2)/U(1)$ coset, proposed by Faddeev as an effective theory of
QCD  at long distances \cite{Fad}.

\bigskip

\hspace*{-0.7cm} {\large\bf Acknowledgment:}
CA acknowledges support from the Austrian
START award project FWF-Y-137-TEC of N.J. Mauser and J S-G support from the
Spanish
MCyT and FEDER projects FPA2002-01161 and BFM2002-03881 and discussions with
L.A. Ferreira.

\end{document}